\documentstyle[12pt]{article}
\begin{document}
\baselineskip=0.7cm
\newcommand{\EQ}{\begin{equation}}
\newcommand{\EN}{\end{equation}}
\newcommand{\EQA}{\begin{eqnarray}}
\newcommand{\EQN}{\end{eqnarray}}
\newcommand{\e}{{\rm e}}
\newcommand{\Sp}{{\rm Sp}}
\renewcommand{\theequation}{\arabic{section}.\arabic{equation}}
\newcommand{\Tr}{{\rm Tr}}
\renewcommand{\thesection}{\arabic{section}.}
\renewcommand{\thesubsection}{\arabic{section}.\arabic{subsection}}
\makeatletter
\def\section{\@startsection{section}{1}{\z@}{-3.5ex plus -1ex minus
 -.2ex}{2.3ex plus .2ex}{\large}}
\def\subsection{\@startsection{subsection}{2}{\z@}{-3.25ex plus -1ex minus
 -.2ex}{1.5ex plus .2ex}{\normalsize\it}}
\makeatother
\def\thefootnote{\fnsymbol{footnote}}
\begin{flushright}
UT-Komaba 92/13\\
November, 1992
\end{flushright}

\vspace{0.5cm}

\setcounter{footnote}{1}
\begin{center}
\Large
 Matrix Models and 2D Critical String Theory\footnote{
Lecture at the workshop on "Basic Problems in String Theory,"
Yukawa Institute for Theoretical Physics, Kyoto, October 19 - 21, 1992.}\\
---  2D Black Hole  by  $c=1$ Matrix Model  ---
\end{center}

\vspace{0.2cm}
\normalsize
 \begin{center}
Tamiaki Yoneya
\footnote
{e-mail address:\\
yoneya@tkyvax.phys.s.u-tokyo.ac.jp and yoneya@tansei.cc.u-tokyo.ac.jp}\\
{\sl Institute of Physics,
University of Tokyo,
Komaba, Tokyo}

\vspace{0.5cm}
\end{center}

In this talk, we first review the possibility of matrix models toward a
nonperturbative
(critical) string theory. We then discuss whether the $c=1$ matrix model
can describe the  black hole solution of 2D critical string theory.   We
show that there exists a class of integral transformations which send the
Virasoro
condition for the tachyon field around the 2D black hole to that around the
linear dilaton vacuum. In particular, we construct an explicit integral formula
wihich describes
a continuous deformation of the linear dilaton vacuum to the black hole
background.

\vspace{0.2cm}
\begin{center}
Contents
\end{center}
\begin{enumerate}
\item Introduction: Basic problems of critical string theory
\item Interpretation of one-dimensional matrix model as a regularized 2D
critical string theory
\item Can the matrix model describe the 2D black hole?
\item Further remarks
\end{enumerate}
The work reported in section 3 is based on a collaboration\cite{JY} with A.
Jevicki.
\vspace{0.5cm}

\section{Introduction: Basic problems of critical string theory}

There are many unsolved questions in string theory.  A basic
difficulty is that the usual continuum formulation (CFT+integration over
moduli)
of string theory is intrinsically perturbative
with respect to string coupling constant.  Namely, we only know the {\it
Feynman rules} without
appropriate {\it derivation} from first principles. This means that we are not
yet able to identify the
dynamical degrees of freedom and the underlying
higher symmetries of the theory. In particular, being
interpreted as a finite theory of quantum gravity, we do not know how to
describe the (global)
{\it dynamics} of spacetime structure in a stringy language. As a consequence,
the most interesting questions of spacetime geometrodynamics,
such as
background independence,
spacetime singularities,
black hole evaporation, and so on, have not been appropriately posed
within the framework of string theory.

One of traditional approaches to these fundamental questions is the
  string field theory.  In this approach, the  fundamental dynamical degrees of
freedom
are supposed to be the
string fields $\psi [x(\sigma),  \ldots]$ as the functionals of
a one-dimensional curve in target space. We try to
construct (effective) action such that it reproduces the string Feynman rules
in perturbation
theory. Although we expect that this should  be possible, there is no guarantee
that
the above string fields provide  a natural language for formulating the
structure of string
theory, in particular, its symmetry structure. For a most recent attempt toward
a
closed string field theory, see ref \cite{Zwiebach}.

Another now popular view is the  renormalization-group approach.
In this case, the  fundamental dynamical degrees of freedom are supposed to be
2D field theories on world sheets and the  renormalization-group fixed points
are interpreted as the classical equations of motion.
This view point seems at least apparently more suitable
to incorporate the dual symmetric structure of the theory, especially, modular
invariance,
than the string field theory. However, it seems much more difficult to
go beyond perturbation theory with respect to string coupling constant.
For a recent discussion of this approach, see ref \cite{Witten1}.

The recent development of  matrix models suggests a new possibility towards
non-perturbative string
theory, at least, in the case of 2D {\it target} spacetime.
It starts from a completely well-defined quantum mechanics.
We can define the matrix model without assuming  string
coupling constant small. In principle, therefore,  it is formulated
nonperturbatively and suggests an entirely new possibility
concerning the question of the true dynamical degrees of freedom
of string theory.  Unfortunately, however, it is not at all clear at present
whether the matrix models provide a
clue to understand the above fundamental questions of geometrodynamics.

In this lecture, after briefly reviewing the connection between the
matrix model and the 2D critical string,
I would like to make some remarks related to the above problems
by focusing to a concrete question,
 {\it How to understand  the 2D black hole critical string solution (i. e.,
$SL(2,R) /U(1)$ WZW
model) within the framework of $c=1$ matrix model?}
\vspace{0.5cm}

\section{Interpretation of $c=1$ matrix model as a 2D critical string theory}
\subsection{2D critical string}

Let us begin from reviewing why the $c=1$ matrix model may be interpreted as
a critical string theory in a 2D target spacetime. The usual continuum world
sheet action
in 2D spacetime with coordinates
$ X^{\mu}=( \phi , x)$,  where $\phi$ is spacelike and $x$
is timelike in Minkowski metric, is \ ($\alpha'=1$)
\EQ
 S={1 \over 8\pi}\int d^2\xi \sqrt{g} (g^{ab}G_{\mu \nu}(X) \partial_aX^{\mu}
\partial_bX^{\nu}
-2R^{(2)}\Phi(X) + 2 T(X)) .
\EN
The local fields $G_{\mu \nu}(x), \Phi (x)$ and $T(x)$ are background fields
corresponding to spacetime metric, dilaton, and tachyon, respectively.
This describes an conformally invariant field theory when
\EQA
 G_{\mu \nu} &=&  \eta _{\mu \nu},   \\
\Phi &=& -\sqrt{2}\phi,   \\
T(X) &=& 0,
\EQN
since the 2D energy momentum tensor,
\EQ
{\cal T}(z)={1 \over 2}((\partial  x)^2-(\partial  \phi)^2)-\sqrt{2}\partial^2
\phi,
\EN
satisfies the usual OPE with central charge $c=26$. This solution is called the
{\it linear dilaton vacuum}.

A problem of this model as a classical solution of critical string theory is
that the perturbative treatment of this theory may not be justified because the
effective string coupling
constant, $g_{{\rm st}}^2 \sim \e^{-2\sqrt{2}\phi}$, is coordinate dependent
owing to the linear dilaton term and becomes
infinitely large as $\phi \rightarrow -\infty$. This requires to regularize the
theory in some way.
A natural regularization is to utilize  the
condensation of the tachyon field $T$\cite{Pol1}\cite{Dasjevicki}.

Consider the renormalization-group fixed-point equation
(in the low-energy approximation)
for the background fields,
\EQ
R_{\mu \nu}- 2\nabla_{\mu}\nabla_{\nu}\Phi + \nabla_{\mu}T\nabla_{\nu}T = 0,
\EN
\EQ
R+4(\nabla \Phi)^2-4\nabla^2 \Phi +(\nabla T)^2+V(T) = 8,
\label{dilatoneq}
\EN
\EQ
-2\nabla^2 T +4\nabla^2 \Phi\nabla T+V'(T) = 0,  \label{tachyoneq}
\EN
 where the tachyon potential is
 \EQ
  V(T)=-2T^2+O(T^3).
\EN
Note that the value $8$ in the rhs of the eq. (\ref{dilatoneq})
due to the dimensionality (2D) of the target spacetime.
It is easy to check that the linear dilaton vacuum is an exact solution of the
fixed-point equation, as it should be.
The linearized tachyon equation ($\sim $ on shell condition) in this background
reduces to
\EQ
(-\partial_x^2 + \partial_{\phi}^2+2\sqrt{2}\partial_{\phi}+2)T=0.
\EN
This allows a static solution
\EQ
T  =b (\phi-\phi_0)\e^{-\sqrt{2}(\phi -\phi_0)}, \quad (\phi_0, b= {\rm
constants})
\EN
Here, $\phi_0$ is an integration constant.  A multiplying constant
$b$ is assumed to be positive.
Adding this term to the world sheet action effectively suppresses the region
where
$\phi\rightarrow -\infty$.
This puts an effective cutoff for the coordinate $\phi$,
\EQ
\phi_0<\phi.
\label{lowercutoff}
\EN

\noindent
Remarks:
(1) This is an {\it infrared} cutoff from the point of view of random surface
in which the tachyon condensation amounts to nonzero
cosmological constant with respecto to world sheet;(2)
The argument is far from rigorous: there is no known exact solution of the
fixed-point equation with nontrivial tachyon field.

Because of the condensation of tachyon, there exists a nonvanishing
contribution for the classical vacuum energy, in contrast to
the usual critical string solutions in $D=26$ or $D=10$ spacetimes.
Actually the total classical energy (i. e., integrated over
$\phi$) is divergent for $\phi \rightarrow
\infty$. To make the theory completely finite, we then have to
introduce another cutoff, say, $\phi < 0$ by adjusting the
integration constant $\phi_0$ appropriately. This cutoff
is ultraviolet in terms of the random surface view point, while
is infrared in terms of the target spacetime picture. The introduction
of an infrared cutoff is natural from the target-space view point,
since the {\it tachyon} is massless in the
linear dilaton vacuum.
\EQ
(-\partial_x^2 +\partial_{\phi}^2){\tilde T}=0, \quad T =
\e^{-\sqrt{2}\phi}{\tilde T}.
\label{lineartachyoneq}
\EN
We are considering the system in a finite box and then take an inifite-volume
limit.

Then, the classical (sphere) free energy,
$F$= [range of x direction]$\times f(\Delta)$,
turns out to be
\EQ
f(\Delta)={1\over 2\pi g_0^2} {\Delta^2 \over \log \Delta}+ \cdots={1\over 2\pi
g_0^2}\mu^2 \log \mu
+\cdots,
\EN
\EQ
T(0)=b\e^{-\sqrt{2}\phi_0}\equiv\Delta, \qquad \phi_0={1 \over \sqrt{2}}\log
\mu.
\EN
This coincides with the famous result \cite{c11}\cite{c12}\cite{c13}\cite{c14}
 of the matrix model if $\Delta$ is identified with the
bare cosmological constant on the world sheet and the bare
string coupling constant $g_0$ is scaled to be proportional to
$1/\beta\sim 1/N$. The double scaling limit is defined as the limit
$N\rightarrow \infty$ with $\beta\mu$ being kept fixed.
It is not difficult, using the continuum approach, to estimate the
genus-one correction $f^{(1)}$ to the free energy.
The result is, using the standard moduli parameter $\tau$,
\EQ
f^{(1)}=-{1 \over 8\pi}\log \mu \int_{{\cal F}} d^2 \tau {1 \over \tau_2^2}=-{1
\over 24}\log \mu,
\EN
and agrees with the matrix model. Note that the integration region
with respect to the moduli parameter is the fundamental region ${\cal F}$.

It is remarkable that
 there is no ultraviolet (from the view point of target spacetime) divergence
in the genus-one free energy
in contrast to local field theory in which case the fundamental
region ${\cal F}$ is replaced by the usual
integration over the proper time $0<\tau_2<\infty$.
There are also many evidences\cite{GrossKlebanov} for the equivalence in the
level of
correlation functions.
{}From these observations, it seems very natural to regard the matrix model as
a toy model which embodies potentially the whole non-perturbative information
of 2D critical string theory.

\subsection{2D black hole solution}

Let us next review the 2D black hole solution in string theory.
The linear dilaton vacuum  as a solution to the fixed point equation
is a special case of the following black hole solution\cite{blackholesolution}
(in the conformal gauge for the
target spacetime metric
$G_{\mu \nu}=\e^{\sigma}\eta_{\mu \nu}$,  $(u, v)$=light-like coordinates),
\EQ
\e^{-2\Phi}=\e^{-\sigma}=uv-C, \qquad  (C={\rm const.}), \qquad T=0.
\label{blackholebackground}
\EN
This reduces to the linear dilaton vacuum in the limit $\phi \rightarrow
\infty$ (or equivalently $C\rightarrow 0$) upon identification,
\EQ
u=\e^{{1 \over \sqrt{2}}(\phi+x)}, \qquad v=-\e^{{1 \over \sqrt{2}}(\phi-x)}.
\label{minkowskicoordinate}
\EN
The exact conformal field theory describing the solution is shown by
Witten\cite{Witten2} to be the
$SL(2,R)/U(1)$ WZW model with $k={9 \over 4}$,
whose action is given by
\EQA
S &=& {-k \over 8\pi} \int {\rm
Tr}[(g^{-1}\partial_{+}g)(g^{-1}\partial_{-}g)]-ik\Gamma_{{\rm WZ}}
\nonumber \\
  & -&{ k\over 4\pi}\int {\rm
Tr}(A_-g^{-1}\partial_{+}g+A_+g^{-1}\partial_{-}g-2A_+A_-+A_+gA_-g^{-1}).
\label{wzwaction}
\EQN
In view of the correspondence of the $c=1$ matrix model with
the linear dilaton vacuum solution, a question arises: Is it possible to
represent the black hole solution in the language of
the matrix model?

\subsection{Discrete states of the $c=1$ model and the black hole}

Since for  $-uv \rightarrow \infty$, the black hole background reduces
to the linear dilaton background, let us first study the nature of the
deformation of the linear dilaton vacuum to the black hole solution. Consider
the
asymptotic expansion of the black hole metric using the coordinates
(\ref{minkowskicoordinate}):
\EQ
d^2s = {du dv \over uv -C}={1\over
2}[(d\phi)^2-(dx)^2+C\e^{-2\sqrt{2}\phi}((d\phi)^2-(dx)^2)+{\rm O}(C^2)].
\EN
Note that the parameter $C$ sets the scale of the black hole mass. We can set
$C=1$  by a scale transformation of the light-like
coordinates.
The first order deformation represented by the  term proportional to
$C$
is equivalent to adding a term,
\EQ
e^{-2\sqrt{2}\phi}((\partial \phi)^2-(\partial x)^2),
\label{deformation}
\EN
to the world sheet action density.
Using the language of CFT, this should be interpreted as a deformation by a
marginal operator. In 2D critical string theories,
there is no higher excited states than the
lowest tachyon modes.
However, it is well known\cite{discretestates} that there are an countably
infinite number of
other physical states with special discrete values of x-energy and
$\phi$-momentum
of the form,
\EQ
{\cal O}_{r,s}=\e^{ipx}\e^{\beta(p)\phi}\times ({\rm Polynomials \, of \,
derivatives of }X^{\mu}).
\EN
The discrete states may be  regarded as the remnants of the physical
excitations
in higher dimensional critical string theories.
The  allowed energies $p$ and momenta $\beta$ of the discrete states are
\EQ
p= {r-s \over \sqrt{2}}, \quad \beta_{\pm}
  ={-2 \pm (r+s) \over \sqrt{2}}
\EN
at the level $n=rs$ with $(r,s)=$integers.
Let us call the discrete operators with $\beta_+$ and $\beta_-$, the
positive and  negative
discrete operators, respectively.

Now we consider in particular the zero-energy ($r=s$) discrete states which
have the following $\phi$-momenta.

\vspace{0.5cm}
\begin{center}
\begin{tabular}{|c|c|c|} \hline
$r$ & $\beta_+$ & $\beta_-$ \\ \hline\hline
$1$ & $0$  	& $-2\sqrt{2}$ \\
$2$ & $\sqrt{2}$ & $-3\sqrt{2}$ \\
$3$ & $2\sqrt{2}$ & $-4\sqrt{2}$ \\
$\vdots$ & $\vdots$ & $\vdots$ \\ \hline
\end{tabular}
\end{center}

\vspace{0.5cm}
\noindent
Comparing with (\ref{deformation}), we see that
the deformation occurring in the black hole solution corresponds to a negative
operator with $\beta_- = -2\sqrt{2}$
at level $n=1$.
Another discrete state  at level $n=1$ with $\beta_+=0$,
$(\partial\phi)^2-(\partial x)^2$, can be easily seen to correspond to a global
rescaling of the coordinate $(x, \phi)\rightarrow (1+\epsilon)(x, \phi)$. This
deformation should be interpreted as a
special case of {\it global} gauge transformations whose gauge parameters grow
indefinitely at infinity.
In general, the positive discrete states with $\beta_+$ are known to generate a
closed symmetry algebra
$W_{1+\infty}^+$, isomorphic to the Lie algebra of area -preserving
diffeomorphisms in two dimensions.
It is an important unsolved question to clarify the nature of this infinite
symmetry structure
in connection with possible higher symmetries underlying the critical string
theory.
On the other hand, from the view point of Liouville theory, it has been
argued\cite{Seiberg} that
the states with $\beta_-$ can not be
represented by local operators.

\section{Can the matrix model describe the 2D black hole?}

Thus, in order to understand the formation of black holes in terms
of the matrix model, it is important to
identify the negative discrete operators.
Before attacking to this
question, let us briefly review how to understand the tachyon and the
discrete states in the $c=1$ matrix model.

\subsection{Fermi fluid picture to the matrix model}

One-dimensional matrix model, after restricting to $U(N)$ singlet
states,
\EQ
L(M,\dot M)={1 \over 2}\Tr[\dot M^2 -V(M)]
\EN
is equivalent
to a non-interacting fermion system,
\EQ
S = \beta \int d\lambda dx \psi^{\dagger}(-{1 \over \beta}
{d\over dx}+{1 \over 2\beta^2} {d^2 \over d\lambda^2}+V(\lambda))\psi,
\EN
\EQ
\int d\lambda  \psi^{\dagger}\psi=N,
\EN
\EQ
\{\psi(\lambda, x), \psi(\lambda', x)\}=\delta (\lambda-\lambda').
\EN
The constant  $\beta^{-1}$ plays the role of $\hbar$.
The following table provides a  dictionary between the two representations.
\vspace{0.5cm}
\[
\begin{array}{ccc}
{\rm [matrix]}  & & {\rm [fermion]} \\
& & \\
\int dx \e^{ipx}\Tr F(M) & \Leftrightarrow & \int dx d\lambda
\e^{ipx}\psi^{\dagger}
(\lambda, x)\psi(\lambda,x)F(\lambda).
\end{array}
\]

\vspace{0.5cm}
\noindent
To extract the tachyon field, one of the shortest ways is to utilize the
fermi-fluid picture in the semi-classical approximation as first
pointed out by Polchinski\cite{Pol2}. This is equivalent with the
collective field method\cite{Dasjevicki}.
 Let $\alpha_{\pm}$ = upper and lower edges of the fermion sea.
Then, by the Bohr-Sommerfeld condition, we have the following bosonized
representation of the fermion bilinears:
\vspace{0.5cm}
\begin{center}
\[
\begin{array}{ccc}
\psi^{\dagger} \psi& \Rightarrow & \beta\int_{\alpha_-}^{\alpha_+}{dp \over
2\pi}={\beta\over 2\pi}(\alpha_+-\alpha_-) \\
{1\over \beta^2}{\partial \psi^{\dagger} \over \partial \lambda}
{\partial \psi \over \partial \lambda} &\Rightarrow &
{\beta\over 2\pi}\int dp \, p^2={\beta \over 6\pi}(\alpha_+^3 +
-\alpha_-^3) \\
\cdots & \cdots & \cdots .
\end{array}
\]
\end{center}

\vspace{0.5cm}
The quantization condition (or more precisely,the Poisson bracket
multiplied by ${i\over \beta^2}$ since the semi-classical
approximation is assumed) for the fields $\alpha_{\pm}$ turns out
to be
\EQ
[\alpha_{\pm}(\lambda), \alpha_{\pm}(\lambda')]=\mp{2\pi i\over
\beta^2}\partial_{\lambda}\delta(\lambda-\lambda'),
\EN
from the above dictionary. On the basis of the above
correspondence, we found that,
in terms of the bose fields $\alpha_{\pm}$, the hamiltonian and the equation of
motion are, respectively,
\EQ
H=\beta\int d\lambda \int_{\alpha_ -(\lambda) }^{\alpha_{+}(\lambda) } dp\,
h(p,\lambda),
\EN
\EQ
{\partial \alpha_{\pm} \over \partial x}=-\partial_{\lambda}
{\partial H \over \partial \alpha_{\pm}},
\EN
where $h(p, \lambda) $ is the one-body hamiltonian
\EQ
h(p,\lambda)={1 \over 2}p^2+V(\lambda).
\EN
Since  the system is symmetric under $p \leftrightarrow -p$,
there is a static classical solution such that
\EQ
\alpha_{\pm}=\pm \alpha_0(\lambda), \qquad h(\alpha_0, \lambda)=0.
\EN
Then the linear perturbation $\tilde\alpha=\alpha_{\pm}\mp\alpha_0$
around the classical solution satisfies the massless Klein-Gordon equation,
\EQ
\tilde\alpha_{\pm}\equiv ({\partial \lambda \over \partial
\sigma})^{-1}\Pi_{\zeta}\mp \partial_{\sigma}\zeta,
\EN
\EQ
(\partial_x^2-\partial_{\sigma}^2)\zeta=0.
\label{collectivetachyoneq}
\EN
\EQ
{\partial \lambda \over \partial \sigma}\equiv \alpha_0(\lambda)
\label{sigmadef}
\EN
This shows that $\zeta$ should be interpreted as the tachyon field $\tilde T$.
Note that (\ref{sigmadef}) defines the spatial coordinate $\sigma$
as a parametrization of the fermi surface.

Now, as is well understood, the scaling limit is governed by the inverse
harmonic potential,
\EQ
V(\lambda)= -{1 \over 2}\lambda^2+\mu.
\EN
The one-body hamiltonian describing the scaling limit thus takes a very simple
form,
\EQ
h=  {1 \over 2}(p^2-\lambda^2)+\mu = {1\over 4}(A_+A_- +A_-A_+)
+\mu,
\EN
where
\EQ
A_{\pm}=p\pm \lambda
\EN
satisfying the commutation relations
\EQ
 [A_+, A_-]=i{2\over  \beta},
\EN
\EQ
 [h, A_{\pm}]= \mp {i\over \beta}A_{\pm}.
\EN
Namely,  $A_{\pm}$  are energy eigenoperators with pure imaginary (Minkowski)
energies (energies become real after Wick rotation to Euclidean metric);
\[
A_+^rA_+^s \sim \e^{(s-r)x}.
\]
Equivalently, these operators generate an infinite dimensional symmetry of the
system
\EQ
[ {i\over \beta}{\partial \over \partial x}-h,\, \e^{-(s-r)x}A_+^rA_-^s ]=0
\EN
The Poisson brackets among these generators form the  $W_{1+\infty}$ algebra,
\EQA
\{H_{J,m}, H_{J',m'}\}&=&(mJ'-m'J)H_{J+J'-1, m+m'}, \qquad (\vert m\vert \le J)
\\
H_{J,m}&\equiv&A_+^{J+m}A_-^{J-m}.
\EQN
These
properties\cite{Avanjevicki}\cite{dasetal}\cite{Moorseiberg}\cite{Minicetal}
 are similar to the discrete states \cite{Witten3}in the linear dilaton vacuum
provided those are identified with the {\it positive} discrete states with the
correspondence $ J={r+s \over 2}, \, m={r-s \over 2}$.

Unfortunately, there is no clear indication of  negative discrete states in the
matrix model.
This problem is perhaps related with the fact that the negative discrete states
can not
be represented by local operators in the language of 2D gravity on the world
sheet.
However, we have seen that the negative operator is essential for understanding
the deformation of the linear dilaton vacuum into the
black hole background.

\subsection{Deformation as canonical transformation}

In the matrix model, the deformations by the positive discrete states
can be regarded as canonical transformations.
Consider to deform the theory by adding the discrete operator to $h$,
\EQ
h\rightarrow h+\epsilon \e^{(r-s)x}H_{J,m}.
\EN
The case $J=1, m=0$, for instance, corresponds to a global rescaling
of spacetime coordinates which is a marginal deformation in the sense of  CFT.
as discussed in the subsection 2.3 above. It is
important to notice that,
since
\EQ
{d \over dx}(\e^{(r-s)x}H_{J,m})=0,
\EN
the deformation by the discrete operator,
 in the language of the matrix model,
 can be regarded as a canonical transformation with
\EQ
F_{J, m}(x, \partial_x)\equiv x \e^{(r-s)x}H_{J,m} \label{generator1}
\EN
being the generating function.  Note that the presence of a factor $x$ in
the generator (\ref{generator1}),
without which the generating function (\ref{generator1}) is a
symmetry generator. From this simple example,
it seems natural to suppose that the deformation by negative discrete states
are
also generated by more complicated (and possibly singular) canonical
transformations.

\subsection{Integral transformations of the tachyon field}

A GKO analysis\cite{Dijkgraafetal} of the $SL(2,R)/U(1)$ WZW model shows that
the exact on-shell
condition for the tachyon in the black hole background takes the following
form,
\EQ
L_ 0^B T(u,v)=T(u,v),
\EN
with
\EQ
L_0^B={1\over k-2}[(1-uv)\partial_u\partial_v
-{1\over 2}(u\partial_u+v\partial_v)]
-{1\over 2k(k-2)}(u\partial_u-v\partial_v)^2.
\label{blackholetachyoneq}
\EN
In the asymptotic limit $u,-v \rightarrow \infty$, this of course reduces to
the
linear dilaton case.
In a previous subsection, we have seen for static
background that the scaling limit of the
$c=1$ matrix model inevitably gives the linearized massless tachyon
equation (\ref{collectivetachyoneq}) (i. e., Klein-Gordon equation).
The assumption of a static classical background for the
bosonized field is natural since the classical black hole background
(\ref{blackholebackground})
has a Killing symmetry under the vector field $u\partial_u-v\partial_v$.
Thus the possibility of describing the
black hole background within the framework of the matrix model
 requires, as a
{\it necessary} condition, that the tachyon operator (\ref{blackholetachyoneq})
around the black hole background should
also be rewritten in the Klein-Gordon form without external fields.

A first step to embed the black hole solution into the matrix model
may then be to find a canonical transformation such that it reduces the
tachyon equation (\ref{blackholetachyoneq}) to the Klein-Gordon form.
Such a possibility\footnote{
For other approaches, see \cite{Ber} \cite{eguchi}. } has in fact been
suggested previously by
Martinec and Shatashvili\cite{Martinec} from a different context. They argued
that
the dual transform of the path-integral representation of the
model (\ref{wzwaction}) is essentially a Liouville theory coupled
with the $c=1$ conformal matter.
It is indeed possible to construct a transformation which sends
(\ref{blackholetachyoneq}) to the form (\ref{collectivetachyoneq}),
on the basis of their observation.
Here we take a different and more
direct approach.

 The general integral representation\cite{Dijkgraafetal} for the
solutions of (\ref{blackholetachyoneq}) is given by
\EQ
T(u,\,v)=\int{dx\over x}(x{A\over B})^{-2i \omega}(AB)^{-1+2i\lambda}
\label{integralrepresentation},
\EN
where
\EQA
A &=& (\sqrt{1-uv}+{u\over x})^{{1\over 2}}\equiv [u(-{1\over x_2 }+
{1\over x})]^{{1\over 2}},
\label{A} \\
B &=& (\sqrt{1-uv}-vx)^{{1\over 2}}\equiv[-v(x-x_1)]^{{1\over 2}}
\label{B}.
\EQN
The on-shell condition is
\EQ
\lambda =\pm {\omega \over 3}.
\EN
There are two independent choices for the integration contours:
two among
\[
C_1=[0,\infty[,\,  C _2=[x_2,0],\, C_3=[x_1, x_2], \,
C_4=]-\infty, x_1].
\]
Take, for instance, the contour $C_1$ and
consider  the region $ u\equiv \sinh {r\over 2}\e^t(>0), \,v\equiv -\sinh
{r\over 2}\e^{-t}(<0)$ outside the horizon.
We then have
\EQ
T_{C_1}(u,\,v)=\e^{-2i\omega}(-uv)^{-{1\over 2}-i\lambda}B(\nu_+, \nu_-)
F(\nu_+,\nu_-;1-2i\lambda;{1\over uv}),
\label{TC1}
\EN
with $\nu_{\pm}={1\over 2}-i(\lambda\pm \omega)$.  The solution (\ref{TC1})
behaves for large $r$ as
\EQ
T_{C_1}(u,\,v)\sim B(\nu_+, \nu_-)\e^{-{r\over 2}}\e^{i\lambda r- 2i\omega t}.
\EN
Other choices give similar expressions in terms of the hypergeometric
functions.

Now, in the integral representation of $T_{C_1}$, let us make the following
change of integration variables,
\EQA
x{A\over B} &=& \e^{-\tau},\\
AB&=& \e^{\sigma}.
\EQN
It is easy to see that in terms of new variables, it takes the form
\EQ
T_{C_1}=\int d\tau d\sigma \, \delta ({ -v\e^{\tau}+u\e^{-\tau} \over 2}
-\sinh \sigma) \e^{-2i\omega \tau -2i\lambda \sigma}.
\label{tc1kernel}
\EN
Thus, $\delta ({-v\e^{\tau}+u\e^{-\tau} \over 2}
-\sinh \sigma)$ plays the role of  kernel for an integral
transformation to the plane-wave solution of the massless
Klein-Gordon equation. Similarly, for the  solution
\EQ
 T_{C_2}(u,\,v)=\e^{-2i\omega}(-uv)^{ -i\omega}B(\nu_+, \overline{\nu_-})
F(\nu_+,\overline{\nu_-};1-2i\omega;  uv ),
\EN
satisfying  the boundary condition near horizon $r \rightarrow 0$,
\EQ
T_{C_2}(u,\,v)\sim B(\nu_+, \overline{\nu_-}) u^{-2i\omega},
\EN
we have, for the same region for $u,\, v$ as above,
\EQ
T_{C_2}=\int d\tau d\sigma \, \delta ({ -v\e^{\tau}+u\e^{-\tau} \over 2}
-\cosh \sigma) \e^{-2i\omega \tau -2i\lambda \sigma}
\label{tc2kernel}.
\EN
Denoting these integration kernels by $M(u,v;\tau, \sigma)$,
we can directly check an operator relation
\EQ
L_0^B  M(u,v;\tau, \sigma) = M(u,v;\tau, \sigma)  L_0^0  ,
\label{transformL0}
\EN
\EQ
L_0^0\equiv -{1 \over 4(k-2)} \partial _{\sigma} ^2+{1 \over 4k}\partial
_{\tau}^2
+{1\over 4(k-2)}.
\EN
Comparing the rhs  of this relation
with (\ref{lineartachyoneq}), we see that the transformed $L_0$ operator gives
the massless tachyon equation $L_0^0=1$ in the
linear dilaton vacuum {\it if and only if} $k={9\over 4}$ after
a trivial renaming of the coordinates.

The above transformations for $T_{C_1}$
can formally be characterized by the following correspondence using  the
conventions of (\ref{transformL0}) ,
\EQA
 \partial_u &\rightarrow& {1\over 2}\partial_{\sigma}{1\over
\cosh\sigma}\e^{-\tau},  \label{transformorg1}\\
 \partial_v &\rightarrow& -{1\over 2}\partial_{\sigma}{1\over \cosh\sigma}\e^{
\tau},  \\
 u &\rightarrow&   (\cosh \sigma \partial_{\tau} +\sinh\sigma
\partial_{\sigma})\partial_{\sigma}^{-1}\e^{\tau},  \\
 v &\rightarrow&   (\cosh \sigma \partial_{\tau} -\sinh\sigma
\partial_{\sigma})
 \partial_{\sigma}^{-1}\e^{-\tau}.
 \label{transformorg2}
\EQN
Here, the lhs are the operators for the black hole background and the rhs
the ones in the linear dilaton vacuum.
For $T_{C_2}$, $\cosh\sigma$ and
$\sinh\sigma$ should be interchanged.

These forms (\ref{transformorg1})$\sim$(\ref{transformorg2}) may be interpreted
as the quantized version of a canonical
transformation in the 4-dimensional phase space consisting of the one-particle
coordinates and momenta in 2D target spacetime of the matrix model, at least
around the
fermi surface.
The functions $\sqrt{2\vert \mu\vert}\sinh \sigma$ or $\sqrt{2\vert
\mu\vert}\cosh \sigma$  appearing in the kernel,
 are then naturally identified
with the  coordinate $\lambda$ representing the matrix eigenvalue whose
functional form in terms of $\sigma$ is defined by (\ref{sigmadef}) with
negative
$\mu$ or positive $\mu$, respectively.
{}From this point of view, however, the transformation is unsatisfactory.
The reason is that
(1) it does not  take a form of
a unitary transformation in the Hilbert space of one-
body problem, namely, as a unitary transformation
for the fermion wave functions; (2) the asymptotic limit $-uv \rightarrow
\infty$ of the
transformation does not reduce to the identity, even though the black hole
background reduces to the linear dilaton vacuum in this limit,
as is seen from (\ref{transformorg1})$\sim$(\ref{transformorg2}).
We conjectured  that  the black hole background can be treated
within the matrix model by making a special canonical
transform. This would require  the
existence of a (at least formally) unitary transformation in the space of
fermion
single-particle space. On the
other hand, our discussion on the deformation of the linear dilaton
vacuum into the black hole background
in the asymptotic region suggests the existence of
a transformation which  reduces to the identity   as we
go far away from the horizon.
We will discuss these problems in a forthcoming paper\cite{JY}.
Below we only briefly discuss the
second problem.

\subsection{A candidate  transformation corresponding to
the black hole deformation}

 We first introduce a
one-parameter family of a new integral kernel $K_a(\tilde u,\tilde v ; \tau,
\sigma)$
 satisfying the following conditions ($a \ne -1$),
 \EQA
 -{1 \over
2}\e^{\tau}[-\e^{\sigma}+(a+1-\partial_{\tau})\partial_{\sigma}^{-1}e^{\sigma}]
K_a &=&
 K_a \tilde u,  \label{10}\\
 -{1 \over
2}\e^{-\tau}[-\e^{\sigma}+(a+1+\partial_{\tau})\partial_{\sigma}^{-1}e^{\sigma}] K_a &=&
 K_a \tilde v, \label{1}
 \EQN
 \EQA
  e^{-\tau - \sigma}\partial_{\sigma}K_a &=& K_a \partial_{\tilde u},  \\
 -e^{ \tau - \sigma}\partial_{\sigma}K_a &=& K_a \partial_{\tilde v}.
 \EQN
Using the  identities ($a \ne -1$),
\EQA
(v\e^{\tau}-\e^{\sigma})z^a&=&(a+1 +\partial_{\tau})\int dz z^a,  \\
(-u\e^{-\tau}-\e^{\sigma})z^a&=&(a+1 -\partial_{\tau})\int dz z^a
\EQN
with $z\equiv {-v\e^{\tau}+u\e^{-\tau}\over 2}-\e^{\sigma}$,
we can easily check that the solution is, for $a\ne -1$,
 \EQ
 K_a=( {-\tilde v\e^{\tau}+\tilde u\e^{-\tau}\over 2}-\e^{\sigma})^a.
 \EN
For the case $a=-1$ in which case the identities are violated,
the solution is $\delta (z)$.
Let us choose the case $a=-2$. Then, the kernel is characterized by
  \EQA
   {1 \over 2}[\e^{\sigma}+\partial_{\tau} \partial_{\sigma}^{-1}e^{\sigma}]
\e^{\tau}K_{-2} &=&
 K_{-2} \tilde u,  \label{30}\\
{1 \over
2}[\e^{\sigma}-\partial_{\tau}\partial_{\sigma}^{-1}e^{\sigma}]\e^{-\tau}
K_{-2} &=&
 K_{-2} \tilde v, \label{3}
 \EQN
 \EQA
  e^{-\tau - \sigma}\partial_{\sigma}K_{-2} &=& K_{-2} \partial_{\tilde u}, \\
 -e^{ \tau - \sigma}\partial_{\sigma}K_{-2} &=& K_{-2} \partial_{\tilde v}.
 \label{newtransform2}
 \EQN
 Note the position of $\e^{\pm \tau}$ in (\ref{30}) and (\ref{3}) comparing
with (\ref{10}) and (\ref{1}).

On the other hand,   the kernel  $M(u,v;\tau, \sigma)\equiv \delta ({-v
\e^{\tau}+u\e^{-\tau} \over 2}-{\sinh \sigma \over 2})$
in   (\ref{tc1kernel}) is characterized by
(\ref{transformorg1})$\sim$(\ref{transformorg2}), namely,
 \EQA
 uM&=&M( \cosh \sigma \partial_{\tau}+\sinh \sigma
\partial_{\sigma})\partial_{\sigma}^{-1}\e^{\tau}  \label{orgtrans1}, \\
 vM&=&M( \cosh \sigma \partial_{\tau}-\sinh \sigma
\partial_{\sigma})\partial_{\sigma}^{-1}\e^{-\tau},
 \EQN
 \EQA
 \partial_u M &=& M   {1 \over 2} \partial_{\sigma}{1 \over \cosh \sigma}
e^{-\tau}.   \\
 \partial_v M &=& -M  {1 \over 2} \partial_{\sigma}{1 \over \cosh \sigma}
e^{\tau}.
 \label{orgtrans2}
 \EQN
We see that the rhs of the asymptotic form of (\ref{orgtrans1})$\sim$
(\ref{orgtrans2}) for large $\sigma$
coincide with the lhs of (\ref{30})$\sim$(\ref{newtransform2})
apart from a factor $\e^{\sigma}$.

Based on this observation, we define a new kernel $G$ by
\EQ
G(u,v;\tilde u, \tilde v)\equiv -{1\over 8\pi^2}\int d\tau d\sigma M(u,v;\tau,
\sigma)e^{\sigma}K_{-2}(\tilde u,\tilde v ; \tau, \sigma) \label{8}.
\EN
This satisfies, in the limit of large $u,-v$ and hence for large  $\tilde u,
-\tilde v$,
 the following properties
 \EQA
uG &\sim& G\tilde u,   \\
vG &\sim& G\tilde v, \\
\partial_u G&\sim& G \partial_{\tilde u}, \\
\partial_v G&\sim& G \partial_{\tilde v}, \label{9}
\EQN
and
\EQ
L^B_0(u, v)G = GL_0(\tilde u, \tilde v),
\EN
where
\EQA
L^B_0(u,v)&=&{1 \over k-2}[(1-uv)\partial_u\partial_v-{1\over
2}(u\partial_u+v\partial_v)-
{1\over 2k}(u\partial_u-v\partial_v)^2], \\
L_0(\tilde u,\tilde v)&=&{1 \over k-2}[ -\tilde u \tilde v\partial_{\tilde
u}\partial_{\tilde v}-{1\over 2}(\tilde u\partial_ {\tilde u}+\tilde
v\partial_{\tilde v})-
{1\over 2k}(\tilde u\partial_{\tilde u}-\tilde v\partial_{\tilde v})^2] .
\EQN
 The transformation not only reduces to the identity asymptotically,
 but also sends the black hole background to the linear dilaton vacuum.
This is checked
by redefining new $\tau$ and $\sigma$ coordinates as
\EQ
\tilde u =\e^{ \sigma +   \tau},  \tilde v =-\e^{ \sigma -  \tau}.
\EN
 Then, we have (compare with (\ref{lineartachyoneq}))
\EQ
L_0(\tilde u,\tilde v)= \e^{-\sigma}L_0^0(\sigma, \tau)\e^{\sigma}.
\EN
The new integral transform enable us to identify the first nontrivial negative
discrete operator
in the bosonized form.
More detailed properties of this transformation will be
discussed in \cite{JY}.

Finally,  we note that the classical form of the
above transformation takes a very simple form.
Let the canonical momentum variables, conjugate to
$u, \,v$ and $\tilde u, \,\tilde v$, be $\Pi_u,\, \Pi_v$ and
$\Pi_{\tilde u},\,\Pi_{\tilde v}$, respectively.  The
classical canonical transformation corresponding to $G$ is then given as
\EQA
u&=&\tilde u+{\tilde v\Pi_v^2 \over (\tilde u\Pi_u+\tilde v\Pi_v)^2},  \\
v&=&\tilde v+{\tilde u\Pi_u^2 \over (\tilde u\Pi_u+\tilde v\Pi_v)^2},  \\
\Pi_{\tilde u} &=& \Pi_u (1-{\Pi_u \Pi_v \over (\tilde u\Pi_u+\tilde
v\Pi_v)^2}),  \\
\Pi_{\tilde v} &=& \Pi_v (1-{\Pi_u \Pi_v \over (\tilde u\Pi_u+\tilde
v\Pi_v)^2}).
\EQN
This of course reduces to the identity transformation in the
asymptotic limit.
The generating function for this canonical transformation is
\EQA
F&=& \tilde u\Pi_u +\tilde v\Pi_v + {\Pi_u\Pi_v \over \tilde u\Pi_u+\tilde
v\Pi_v } \label{classicalgeneratingfunction} \\
&\equiv &\tilde u\Pi_u +\tilde v\Pi_v +\tilde F
\EQN
by which the canonical variables are related as
${\partial F \over \partial \Pi_u}=u, {\partial F \over \partial \tilde
u}=\Pi_{\tilde u}, \ldots$.  We can easily check that the classical particle
hamiltonian
satisfies $(1-uv)\Pi_u\Pi_v=-\tilde u\tilde v\Pi_{\tilde u}\Pi_{\tilde v}$.
We may also try to construct the quantum transformation in the
single fermion space directly by  generalizing the classical form
(\ref{classicalgeneratingfunction}).

\section{Further remarks}

There are many questions left. The integral transformations for
the linearized tachyon field  discussed above
indeed describe how to choose the
variables in the phase space, at least,  near the fermi surface, and
suggest the feasibility of our conjecture that
the  black hole solution can be described by
canonical transformations in the matrix model.
In particular, our transformation $G$ may be used
to directly obtain the S-matrix around the black hole
from the matrix model,
since it satisfies the desirable asymptotic properties.
The foregoing discussion is not, however, sufficient to decide whether  the
transformations should be generalized to the whole single-particle
phase space of the free fermion. In connection with this,
possible relations of our picture to the more conventional
approaches as discussed in the first part of this lecture
must also be clarified.

Let us  finally make a comment
related to the question of the correct
dynamical degrees of freedom in string theory.
The fermi-fluid picture  on how the tachyon and the discrete states
appear  in the matrix model shows that the phase space of the
matrix eigenvalue describes both the target space and the
field-degrees of freedom in a unified manner. The fermi surface
is nothing but a space-like surface of the target spacetime,
and the small fluctuation of the surface is interpreted as
the propagation of the tachyon field. A canonical transformation
connecting  different backgrounds, in general,
mixes the target space and the field degrees of freedom.
In this sense, there is a {\it relativity} with respect to  the separation
between the spacetime and the field degrees of freedom.
Clear separation between them is only meaningful in
the weak coupling regime where $\beta\mu$ is large.
This suggests that the correct framework for formulating
short-distance dynamics of string theory as  quantum gravity
might be in some enlarged phase space containing target spacetimes as
 particular sections. As a first step toward the goal, we should be
 able to unify the case $c<1$ and $c=1$ in such a framework.

\vspace{0.5cm}
\noindent
{\large Acknowledgements}\\
I am grateful to A. Jevicki for collaboration and to the members
of my laboratory, especially, K. Hamada, H. Ishikawa,
 M. Kato and Y. Kazama for conversations on 2D strings and black holes. This
work is partialy supported by  Grant-in Aid for Scientific Research on Priority
Areas ``Infinite
Analysis" (04245208) and Grant-in Aid for Scientific Research (04640283).

\vspace{0.7cm}
\noindent
{\it Note added}\\
After the workshop, we found   two preprints (in hepth) \cite{Das}
\cite{Dharetal}
in which  a similar integral
 transform as ours, (\ref{tc1kernel}) or (\ref{tc2kernel}), is
 discussed independently.

\end{document}